\documentclass[10pt]{article}
\usepackage{amssymb, amsmath}
\input epsf.tex

\let\badcite=\cite
\def\cite{~\badcite}

\def\slashchar#1{\setbox0=\hbox{$#1$}           
   \dimen0=\wd0                                 
   \setbox1=\hbox{/} \dimen1=\wd1               
   \ifdim\dimen0>\dimen1                        
      \rlap{\hbox to \dimen0{\hfil/\hfil}}      
      #1                                        
   \else                                        
      \rlap{\hbox to \dimen1{\hfil$#1$\hfil}}   
      /                                         
   \fi}
    \def\slashword#1{\setbox0=\hbox{$#1$}        
  \dimen0=\wd0                                   
   \setbox1=\hbox{/} \dimen1=\wd1                
   \ifdim\dimen0>\dimen1                         
      \rlap{\hbox to \dimen0{\hfil\bf---\hfil}} %
      #1                                         %
   \else                                         
      \rlap{\hbox to \dimen1{\hfil$#1$\hfil}}    
      /                                          
    \fi}                                         %

\catcode`@=11
\newdimen\vbigd@men                             

\def\vbig#1#2{{\vbigd@men=#2\divide\vbigd@men by 2%
   \hbox{$\left#1\vbox to \vbigd@men{}\right.\n@space$}}}
\catcode`@=12

\catcode`@=11
\def\citenum#1{\csname b@#1\endcsname}
\catcode`@=12

\begin{document}
\begin{titlepage}

\begin{flushright}
{SCUPHY-TH-06005}\\
{SCNU-07001}\\
\end{flushright}

\bigskip\bigskip

\begin{center}{\Large\bf\boldmath
Elimination of IR/UV via Gravity  in Noncommutative Field Theory}
\end{center}
\bigskip
\centerline{\bf N. Kersting}
\centerline{{\it Physics Department, Sichuan University, P.R. China 610065}}
\centerline{\bf J. Yan}
\centerline{{\it Physics Department, Sichuan Normal University, P.R. China 610066}}
\bigskip

\begin{abstract}

Models of particle physics with Noncommutative Geometry (NCG) generally suffer from a manifestly non-Wilsonian coupling of infrared and ultraviolet degrees of freedom known as the "IR/UV Problem" which would tend to compromise their phenomenological relevance. In this Letter we explicitly show how one may remedy this by coupling NCG to gravity.
In the simplest scenario the Lagrangian gets multiplied by a nonconstant background metric; in $\phi-4$ theory the theorem that $\int d^4 x \phi \star \phi = \int d^4 x \phi^2$ is no longer true and the field propagator gets modified by a factor which depends on both NCG and the variation of the metric. A suitable limit of this factor as the propagating momentum gets asymptotically large then eradicates the IR/UV problem.
 With gravity and NCG coupled to each other, one might expect anti-symmetric components to arise in the metric.
Cosmological implications of such are subsequently discussed.

\bigskip

\end{abstract}

\newpage
\pagestyle{empty}

\end{titlepage}


\section{Introduction}

Theories of spacetime with noncommutative geometry (NCG) have developed over the years to the point where one can write down a NCG Standard Model of particle physics  \cite{ncsm1, ncsm2}
and derive various phenomenological bounds\cite{pheno1, pheno2, pheno3,pheno4,pheno5} on the parameters of NCG.
This merely requires that ordinary products between fields be replaced by Moyal "star" products,
\begin{equation}\label{starp}
    f(x) ~ g(x) \longrightarrow f(x) \star g(x) \equiv
    f(x) e^{\frac{i}{2}\overleftarrow{\partial}_\mu\theta^{\mu\nu}
    \overrightarrow{\partial}_\nu} g(x)
\end{equation}
where $\theta^{\mu\nu}$ is an antisymmetric object (usually assumed independent of spacetime) parameterizing the strength of NCG,
and that the gauge symmetries $SU(3) \times SU(2) \times U(1)$ of the SM be realized in their enveloping algebrae \cite{gauge}.
Yet such theories  all overlook the so-called "IR/UV Problem"\cite{iruv}, an inconsistency in the way NCG fits into the renormalization procedure. One could argue this is not a fundamental problem because NCG is a low-energy remnant of string theory\cite{string}, which does not have an IR/UV problem. But lack of an explicit way to deal with IR/UV is nonetheless unsatisfactory.

In this work we show how the IR/UV problem might be nullified by simply including a non-constant metric into the theory. We find that if there is a certain intimate relationship between NCG and gravity at high energies, the IR/UV dilemma can be avoided. Assuming that NCG and gravity are coupled theories, we then investigate possible effects of NCG on gravity at cosmological scales, deriving novel solutions to the Friedman equation.

\section{Gravity in the Propagator}

In flat space ($\sqrt{-g} = 1$) it is simple to derive the fact that
\begin{equation}\label{ident}
\int d^4 x (\phi \star \psi) \equiv \int d^4 x
(\phi e^{^ \frac{i}{2}\overleftarrow{\partial}_\mu \theta^{\mu \nu}
\overrightarrow{\partial}_\nu}\psi) = \int d^4 x (\phi \psi)
\end{equation}
for any two fields $\phi$ and $\psi$. This hinges on the identity
\begin{eqnarray}
\phi[\overleftarrow{\partial}_\mu \theta^{\mu \nu}
\overrightarrow{\partial}_\nu]^n \psi
   = \partial_\mu (\phi [\overleftarrow{\partial}_\rho \theta^{\rho \sigma}
\overrightarrow{\partial}_\sigma]^{n-1} \theta^{\mu\nu}
 \partial_\nu \psi )
 \end{eqnarray}
 for $n \ge 1$; hence the exponential in (\ref{ident}) reduces to unity
 plus an infinite sum of total derivatives which vanish (assuming suitable field boundary conditions) upon integration over all
 space.
 Since any field bi-product is thereby unaffected by NCG, propagators in a noncommutative theory are identical to those in the ordinary one.

 However in curved space this no longer holds.
 Assuming first for simplicity that
 the star product only couples matter and does not apply to the metric, a scenario which we designate \textit{minimal NCG coupling}, a typical term in the expansion of
 $\sqrt{-g} ~ \phi \star \psi$ looks like
\begin{eqnarray} \nonumber
 \sqrt{-g}~\{ \phi[\overleftarrow{\partial}_\mu \theta^{\mu \nu}
\overrightarrow{\partial}_\nu]^n \psi \}
& =&
 \partial_\mu (\sqrt{-g}~ \{ \phi [\overleftarrow{\partial}_\rho \theta^{\rho \sigma}
\overrightarrow{\partial}_\sigma]^{n-1} \theta^{\mu\nu}
 \partial_\nu \psi \} ) \\ \nonumber
 && - \sqrt{-g}~ \{ \phi [\overleftarrow{\partial}_\rho \theta^{\rho \sigma}
\overrightarrow{\partial}_\sigma]^{n-1} \theta^{\mu\nu}
 \partial_\mu \partial_\nu \psi \} \\
 &&  - (\partial_\mu \sqrt{-g} ) \{ \phi [\overleftarrow{\partial}_\rho \theta^{\rho \sigma}\overrightarrow{\partial}_\sigma]^{n-1} \theta^{\mu\nu}\partial_\nu \psi \}
\end{eqnarray}
which, upon integration over all space, becomes
\begin{equation}\label{gident}
 \int d^4 x \sqrt{-g}~ \{ \phi[\overleftarrow{\partial}_\mu \theta^{\mu \nu}
\overrightarrow{\partial}_\nu]^n \psi \} =
 -\int d^4 x (\partial_\mu \sqrt{-g} ) \{ \phi [\overleftarrow{\partial}_\rho \theta^{\rho \sigma}\overrightarrow{\partial}_\sigma]^{n-1} \theta^{\mu\nu}\partial_\nu \psi \}
\end{equation}
Specializing to the case $\psi = \phi$ then gives
\begin{eqnarray} \nonumber \label{biprod}
 \int d^4 x \sqrt{-g} (\phi \star \phi)
&=&
 \int d^4 x \sqrt{-g}~\phi^2 \\ \nonumber
 && -   \left[ \begin{array}{c}
      (\partial_\mu \sqrt{-g} ) \phi \theta^{\mu \nu} \partial_\nu \phi \\
      + \frac{1}{2!} (\partial_\mu \sqrt{-g} ) \{ \phi [\overleftarrow{\partial}_\rho \theta^{\rho \sigma}\overrightarrow{\partial}_\sigma]\theta^{\mu\nu}\partial_\nu \phi \} \\ \nonumber
      + \frac{1}{3!} (\partial_\mu \sqrt{-g} ) \{ \phi [\overleftarrow{\partial}_\rho \theta^{\rho \sigma}\overrightarrow{\partial}_\sigma]^2 \theta^{\mu\nu}\partial_\nu \phi \} \\
      +...
    \end{array} \right]  \\
  &=&
   \int d^4 x  \sqrt{-g}~\phi^2 - (\partial_\mu \sqrt{-g} ) \{ \phi
  \left[ \frac{e^{^ {\frac{i}{2}\overleftarrow{\partial}
  \theta \overrightarrow{\partial}}} - 1}
  {\frac{i}{2}\overleftarrow{\partial}
  \theta \overrightarrow{\partial}} \right] \theta^{\mu\nu}\partial_\nu \phi \}
\end{eqnarray}
in a more formal notation. Since the star-product of two fields is now no longer equal to the ordinary product, we expect the form of the scalar propagator to change.

For definiteness, consider the NC scalar theory defined by the action
\begin{equation}\label{scalarlag}
    S = \int d^4 x \sqrt{-g}(\frac{1}{2}{\partial}_\mu \phi \star {\partial}^\mu \phi + \frac{1}{2} m^2 \phi \star \phi
    + \frac{\lambda}{4!}\phi \star \phi \star \phi \star \phi )
\end{equation}
Concentrating on the kinetic terms and employing (\ref{biprod}),
\begin{eqnarray} \nonumber
    \int d^4 x \sqrt{-g}(\frac{1}{2}{\partial}_\mu \phi \star {\partial}^\mu \phi + \frac{1}{2} m^2 \phi \star \phi) =
    \int d^4 x \sqrt{-g}(\frac{1}{2}(\partial \phi)^2 - \frac{1}{2} m^2 \phi^2 )
    \\
    + ~ \theta^{\mu \nu} \frac{\partial_\mu \sqrt{-g}}{2}
    \{ \partial^\rho \phi
  \left[ \frac{e^{^ {\frac{i}{2}\overleftarrow{\partial}
  \theta \overrightarrow{\partial}}} - 1}
  {\frac{i}{2}\overleftarrow{\partial}
  \theta \overrightarrow{\partial}} \right]\partial_\nu \partial_\rho \phi \}
  +
  m^2 \theta^{\mu \nu} \frac{\partial_\mu \sqrt{-g}}{2}
 \{ \phi
  \left[ \frac{e^{^{\frac{i}{2}\overleftarrow{\partial}
  \theta \overrightarrow{\partial}}} - 1}
  {\frac{i}{2}\overleftarrow{\partial}
  \theta \overrightarrow{\partial}} \right]\partial_\nu \phi \}
  \end{eqnarray}
from which we derive a modified propagator
\begin{eqnarray}\label{prop}
D(p) &=& \frac{\sqrt{-g}}{(p^2 + m^2)\Delta}\\
\Delta &\equiv & 1-(\partial_\mu \sqrt{-g})p_\nu \theta^{\mu \nu}
\end{eqnarray}
which is equal to the ordinary one (in a background metric) divided by a correction factor $\Delta$ which we have left in a mixed momentum-position space notation. This factor departs from unity only in the case where space is both curved and noncommutative. Had we in fact considered a full coupling of NCG to the theory with
$\sqrt{-g} \to \sqrt{-g} ~\star $ in (\ref{scalarlag}), it is not difficult to see that $\Delta$ would receive additional corrections of the form
$p_{\rho_1} p_{\rho_2}...p_{\rho_n} \theta^{\rho_1 \sigma_1}...\theta^{\rho_n \sigma_n}
\partial_{\sigma_1}\partial_{\sigma_2}...\partial_{\sigma_n} \sqrt{-g}$; this
however does not affect our main results in the following.

\section{Avoiding IR/UV}

The classic demonstration\cite{iruv}  of IR/UV coupling in NCG is the computation of the 1PI correction to the two-point function of the scalar theory (\ref{scalarlag}) in flat space.
At lowest order this is just $\Gamma^{(2)}_0 = p^2 + m^2$, while the one loop corrections are
\begin{eqnarray}\label{2point}
  \Gamma^{(2)}_{1} = \frac{\lambda^2}{6 (2 \pi)^2}
    \int d^4 k \frac{2+ e^{ik_\mu \theta^{\mu \nu} p_\nu}}{k^2 + m^2} \\
\end{eqnarray}
The one loop 1PI quadratic effective action is then
\begin{eqnarray}
  S^{(2)}_{1PI} = \int d^4 p \frac{1}{2}(p^2 + m^2
  +\frac{\lambda^2 }{96 \pi^2 }{\Lambda^2}_{eff}-
  \frac{m^2 \lambda^2}{96 \pi^2}ln({\Lambda^2}_{eff}/m^2))
\end{eqnarray}
where ${\Lambda^2}_{eff} \equiv (1/\Lambda^2 +
p_\mu \theta^{\mu \rho} \theta_{\rho}^{\cdot ~ \nu} p_\nu)^{-1}$, $\Lambda$ being the UV cutoff of the theory. If one takes $\theta \to 0$ in this expression then
$\Lambda \to {\Lambda}_{eff} $ and the theory acquires the usual quadratic and logarithmic divergences as $\Lambda \to \infty$. However if we first take $\Lambda \to \infty$ then we get ${\Lambda}_{eff} \to (p_\mu \theta^{\mu \rho} \theta_{\rho}^{\cdot ~ \nu} p_\nu)^{-1/2}$ and   the UV divergences of the theory disappear, replaced instead by an IR divergence as $p \to 0$. This noncommutivity of the limits $p \to 0$ and $\Lambda \to \infty$ appears to be a disease of the theory.

But now in curved space we must use the altered propagators as in (\ref{prop}) and the 1PI correction is now
\begin{eqnarray}
  \Gamma'^{(2)}_{1} = \frac{\lambda^2}{6 (2 \pi)^2}
    \int d^4 k \frac{2+ e^{ik_\mu \theta^{\mu \nu} p_\nu}}{(k^2 + m^2)\Delta} \\
    \nonumber
\end{eqnarray}
Evidently we can fix the IR/UV problem by requiring that this reduces to the ordinary theory in the high energy limit, i.e.
\begin{eqnarray}
 \frac{2+ e^{ik_\mu \theta^{\mu \nu} p_\nu}}{(k^2 + m^2)\Delta}  ~~~
 \xrightarrow{k \to \infty}  ~~~ \frac{3}{(k^2 + m^2)}  \\ \nonumber
\end{eqnarray}
which implies
\begin{eqnarray}\label{condition}
 \partial_\mu \sqrt{-g} ~~~
 \xrightarrow{k \to \infty} ~~~ \frac{1 - e^{ik_\mu \theta^{\mu \nu} p_\nu}}
 { k_\nu \theta^{\mu \nu}}
\end{eqnarray}
Now the IR/UV problem is manifestly fixed,
\begin{equation}
\frac{\lambda^2}{6 (2 \pi)^2}
    \int d^4 k \frac{2+ e^{ik_\mu \theta^{\mu \nu} p_\nu}}{(k^2 + m^2)\Delta}
  ~~~
 \xrightarrow[k \to \infty]{p \to 0} ~~~  \frac{\lambda^2}{6 (2 \pi)^2}
    \int d^4 k \frac{3}{(k^2 + m^2)}
\end{equation}
 as both IR ($p \to 0$) and UV ($k \to \infty$) limits produce the same result.
 Though curved spacetime and NCG  prima facie would appear to have separate origins,
the physical meaning of (\ref{condition}) is that at high energies gravity and NCG must become dependent on each other: the spatial variation of the metric tends to zero at high energies but in a very nontrivial way.

\section{Cosmology}
In the previous section we saw that to solve the IR/UV problem gravity and NCG must share an intimate relationship at short distances; but at long distance (\ref{condition}) does not assist us. If there is some effect of NCG on gravity
at cosmological scales it must be much smaller than the usual contributions to the metric at these scales from, e.g. dust and dark matter/energy, so as to have escaped detection thus far. On the other hand if we write the metric as
\begin{equation}\label{met}
    g_{\mu \nu} = g_{\{\mu \nu\}} + g_{[\mu \nu]}
\end{equation}
then it is possible that NCG contributes or is solely responsible for the antisymmetric piece $g_{[\mu \nu]}$, since constraints on this are much weaker\cite{asym}.
Let us define a metric
\begin{equation}\label{ametric}
   g_{\mu \nu} =  \left(
      \begin{array}{cccc}
        1 & s & 0 & 0 \\
        -s  & -R^2 & f & 0 \\
        0 & -f & -R^2 r^2 & 0 \\
        0 & 0 & 0 & -R^2 r^2 sin^2 \theta \\
      \end{array}
    \right)
\end{equation}
in spherical coordinates. Here $s$ and $f$ are functions of space and time which are presumed to encode the effects of NCG, possibly breaking isotropy and homogeneity. The case where these are both nonzero is rather complicated; the case where $s=0$ but $f \ne 0$ has been treated elsewhere\cite{moffat}, with the result that both galaxy rotation curves and inflation characteristics can be accounted for providing $f$ satisfies suitable constraints. We have worked out the opposite case where $f=0$ but $s \ne 0$, as we detail below.

Let us  begin with the following action:
\begin{equation}
S=\int {d^{4}x}\sqrt{-g}
[L_{G}+2k(L_{M}+L_{\phi })]
\end{equation}
with the metric as defined in (\ref{ametric}) (but with s = constant for simplicity), $\ L_{G}=\ R$(scalar curvature), $\ L_{M}$ a matter term, and $k$ a coupling constant($=8\pi G$). The scalar piece $L_{\phi }$ is
\begin{equation}
     L_{\phi }=\frac{1}{2}[g^{\mu \nu }\frac{%
\partial \phi }{\partial x^{\mu }}\star \frac{\partial \phi }{\partial x^{v}}%
-m^{2}\phi \star \phi +u_{0}\phi \star \phi \star \phi \star \phi ]
\end{equation}

We assume the energy-momentum tensor of matter fields have a perfect fluid form
\begin{equation}
\ T_{\mu \nu }(M)=(p+\rho )U_{_{^{\mu
}}}U_{\nu }+pg_{_{^{\mu \upsilon }}}
\end{equation}
with $\rho, p$ being matter/pressure density respectively, which gives the energy-momentum tensor of the scalar field as
\begin{equation}
\ \ T_{\mu \nu }(\phi )=\frac{\partial \phi }{\partial x_{\mu }}.\frac{%
\partial \phi }{\partial x_{\nu }}-\frac{1}{2}g_{\mu \nu }[g^{\rho \sigma }%
\frac{\partial \phi }{\partial x_{\rho }}.\frac{\partial \phi }{\partial
x_{\sigma }}-m^{2}\phi ^{2}+
u_0(\phi ^{4}-\phi ^{2}\partial _{\mu }\partial _{\rho }\phi\partial
_{\nu }\partial _{\sigma }\phi\theta ^{\mu \nu }\theta ^{\rho \sigma })]
\end{equation}
to lowest order in $\theta$.

The field equations are then
\begin{eqnarray}
 R_{\mu \nu }-\frac{1}{2}g_{\mu
v}R  &=&-kT_{\mu \nu }(M,\phi )  \\
 \frac{\partial }{\partial x^{u}}(\sqrt{-g}%
g^{\mu \nu }\frac{\partial \phi }{\partial x^{\nu }})  &=&
\frac{1}{2}\sqrt{-g}%
\frac{\partial U_{eff}}{\partial \phi }
\end{eqnarray}
where we have defined the effective potential
\begin{equation}
   U_{eff}\ \equiv u_{0}\phi ^{4}-u_{0}\phi ^{2}\partial _{\mu
}\partial _{\rho }\phi \partial _{\nu }\partial _{\sigma }\phi \theta ^{\mu
\nu }\theta ^{\rho \sigma }-m^{2}\phi ^{2}+\frac{m^{2}}{2}\partial _{\mu
}\partial _{\rho }\phi \partial _{\nu }\partial _{\sigma }\phi \theta ^{\mu
\nu }\theta ^{\rho \sigma }
\end{equation}

We also have the conservation law of the energy-momentum tensor
\begin{equation}
T_{;\nu }^{\mu \nu
}(M,\phi )=0
\end{equation}
For simplicity we assume all components of the NCG parameter $|\theta|~\approx ~ s$;  then for $s<<1,$  the above equations give (in the following
we use a dot for time derivation and a subscript "r" for spatial derivation)

\begin{eqnarray}
  3\frac{\ddot{R}}{R}& =& -\frac{k}{2}%
(1+3\gamma )\rho -k(\dot{\phi}^{2}+\frac{1}{2}U_{eff})+\frac{ks^{2}}{%
2R^{2}}(\dot{\phi}^{2}+U_{eff})\\
  \frac{\ddot{R}}{R}+2(\frac{\dot{%
R}}{R})^{2}&=&\frac{k}{2}(1-\gamma )\rho -\frac{k}{2}U_{eff}+\frac{%
ks^{2}}{2R^{2}}(\dot{\phi}^{2}+U_{eff}) \\
  {(R^{3}\dot{\phi})}^{\prime }&= &\frac{R^{3}}{2}\frac{\partial U_{eff}}{%
\partial \phi } \\
  3(1+\gamma )\frac{\dot{R}}{R}\rho +\dot{\rho} &=& 0
\end{eqnarray}
where we have used the state equation $p=\gamma \rho $.
The Friedman equation is now
\begin{equation}
H^{2} =\frac{k\rho _{eff}}{3},\
\end{equation}
with $H=\frac{\overset{\cdot }{R\ }}{R}$ the Hubble parameter and the effective energy
density
\begin{equation}
\rho _{eff}=\rho +\rho _{\phi }+\rho _{\phi ,s}
\end{equation}
with
\begin{eqnarray}
  \rho &=& \rho _{0}R^{-3(1+\gamma )} \\
  \rho _{\phi } &=& \frac{1}{2}(%
\dot{\phi}^{2}-u_{0}\phi ^{4}+m^{2}\phi ^{2}) \\
  \rho _{\phi ,s}&=&\frac{s^{2}}{2R^{2}}(\dot{\phi}%
^{2}+u_{0}\phi ^{4}-m^{2}\phi ^{2}+u_{0}s^{2}\dot{\phi}^{2}\phi ^{2}\phi
_{r}^{2}-\frac{m^{2}}{2}s^{2}\dot{\phi}^{2}\phi _{r}^{2})
  -\frac{s^{2}}{2}(u_{0}\dot{\phi}^{2}\phi
^{2}\phi _{r}^{2}-\frac{m^{2}}{2}\dot{\phi}^{2}\phi _{r}^{2})
\end{eqnarray}
Ratios of these give
\begin{eqnarray}
  \frac{\rho _{\phi ,s}}{\rho }& =& \frac{s^{2}}{2\rho _{0}}%
R^{-5-3\gamma }[(\dot{\phi}^{2}+u_{0}\phi ^{4}-m^{2}\phi ^{2})+(u_{0}\dot{%
\phi}^{2}\phi ^{2}\phi _{r}^{2}-\frac{m^{2}}{2}\dot{\phi}^{2}\phi
_{r}^{2})s^{2}] \\
 &&-\frac{s^{2}}{2\rho _{0}}R^{3(1+\gamma
)}(u_{0}\dot{\phi}^{2}\phi ^{2}\phi _{r}^{2}-\frac{m^{2}}{2}\dot{\phi}%
^{2}\phi _{r}^{2}) \\
 \frac{\rho _{\phi ,s}}{\rho _{\phi }} &=& 2\rho
_{0}(\frac{\rho _{\phi ,s}}{\rho })R^{-3-3\gamma }[(\dot{\phi}^{2}-u_{0}\phi
^{4}+m^{2}\phi ^{2})]^{-1}
\end{eqnarray}
When $\gamma =-1,$we obtain the following cosmological solutions to the field
equations
\begin{eqnarray}
  R &=& R_{0}e^{-\alpha t}  ~~ ,~~~~ \phi
=\phi _{0} ~~ ,~~~~~   \rho =\rho _{0}  \\
3 \alpha ^{2}&=&k\rho _{0}-\frac{k}{2}(u_{0}\phi
_{0}^{4}-m^{2}\phi _{0}^{2}) \\
  \frac{\partial U_{eff}}{\partial \phi } &=& 0
\end{eqnarray}
From these solutions we have

\begin{eqnarray}
 \frac{\rho _{\phi ,s}}{\rho
} &\backsim &  \frac{m^{4}s^{2}}{8\rho _{0}u_{0}R_{0}^{2}e^{2}} \\
  \frac{\rho _{\phi ,s}}{%
\rho _{\phi }} &\backsim & \frac{s^{2}}{R_{0}^{2}e^{2}}
\end{eqnarray}
where $\ t=\frac{1}{\sqrt{3k}}[(\rho _{0}+(m^{4}/8u_{0})]^{-1/2}.$
These provide novel contributions to density fluctuations which might
be observable in the power spectrum of the cosmic microwave background\cite{cmbrwork}.

\section{Conclusions}

In the foregoing we have seen that one can avoid the IR/UV problem in the two-point function in noncommutative $\phi-4$ theory by including the effects of a background metric. The price one pays is that this metric becomes coupled to NCG at high energies in a nontrivial way. The same mechanism should work to remove IR/UV features in higher correlation functions, since every propagator (\ref{prop}) is designed to cancel the
corresponding NC phase factor (as in (\ref{2point})) which arises from each vertex.

We have not carried out this analysis for fermionic theories, in particular NCQED, to which a great deal of phenomenological work is devoted, though we are confident a similar mechanism is possible there. In such theories there would seem to be more parametric freedom since now the propagator will receive corrections from contractions such as $\partial^\mu (\sqrt{-g})p_\mu p_{\rho}\gamma_\nu \theta^{\nu \rho}$ and
$\partial_\mu (\sqrt{-g}~ g^{[\rho \sigma]})\gamma_\rho p_\sigma p_\nu \theta^{\mu \nu}$ in which the antisymmetric components of the metric participate.\footnote{Note that
the antisymmetric components of the metric $g^{[\rho \sigma]}$ do not appear either in the minimally-coupled or the fully NC $\phi$-4 theory.}

One could also embellish the theory by making $\theta$ itself space-time dependent, or make the full metric (\ref{met}) apply to the star-product as well, i.e.
$\phi \star \phi = \phi ~ exp(\frac{i}{2}\overleftarrow{\partial^\mu}
  g_{\mu \rho}\theta^{\rho \sigma} g_{\sigma \nu} \overrightarrow{\partial^\nu})        \phi$. There are therefore a number of ways gravity could couple to NCG in a way
  which nullifies IR/UV. Though Phenomenologists have largely ignored IR/UV complications in their work thus far, they could for completeness now choose to introduce a non-constant metric, as in our minimal coupling scheme, into their theories. This would not change any of their results but would at least alleviate doubts of the internal consistency of NCG phenomenology.

Thus while the results of this work are unlikely to be testable at high energies, there may be imprints on cosmology: if gravity and NCG are indeed part of the same phenomenon, there may be antisymmetric components of the metric. While \cite{moffat} had done analysis of this possibility for space-space components of the metric, we have begun to do so for space-time components.

\end{document}